# The Strength of the Weak: The Uncertainty Principle and More-Direct Access to the Wave Function through Weak Measurements

David Geelan
Griffith University
d.geelan@griffith.edu.au

Please note, this is not an original research article, it is a brief review of some recent work, published only for interest and as a simple introduction to the concept of weak measurements.

**Abstract**

Recent empirical work in the field of 'weak measurements' has yielded novel ways of more directly accessing and exploring the quantum wavefunction. Measuring either position or momentum for a photon in a 'weak' manner yields a wide range of possible values for the measurement, and can be done in such a way as to only minimally effect the wavefunction rather than to collapse it to a specific precise value. Measuring the other complementary variable (position or momentum) precisely at a later time ('post-selection') and averaging the weak measurements can yield information about the wavefunction that is not directly experimentally obtainable using other methods. This paper discusses two recent papers on weak measurement in the context of the uncertainty principle more broadly, and considers some possibilities for further research.

**Introduction**

Since it was outlined by Werner Heisenberg in 1927, the 'uncertainty principle' has been a foundational part of quantum mechanics. This paper considers the uncertainty relations and discusses 'weak measurements': measurements that give approximate results for the relevant observable such that the wavefunction is not appreciably effected. Averaging a large number of weak measurements of one observable followed by 'post-selection' using strong measurement of another, complementary observable, can yield intriguing information about the particle wavefunction itself. The weak measurement approach was first proposed over 20 years ago, but recent experimental work has explored it in interesting new ways.

*The Uncertainty Principle*

Werner Heisenberg outlined a set of relations for the 'indeterminacy' of the measurement of certain entities in his 1927 paper "Über den anschaulichen Inhalt der quantentheoretischen Kinematik and Mechanik". The title is translated as "The physical content of quantum kinematics and mechanics" in Wheeler and Zurek (1983), however Hilgevoord and Uffink (2011) discuss the translation of the term 'anschaulich' as follows:

> ...the term *anschaulich* is particularly notable. Apparently, it is one of those German words that defy an unambiguous translation into other languages. Heisenberg's title is translated as "*On the physical content …*" by Wheeler and Zurek (1983). His collected works (Heisenberg, 1984) translate it as "*On the perceptible content …*", while Cassidy's biography of Heisenberg (Cassidy, 1992), refers to the paper as "*On the perceptual content …*". Literally, the closest translation of the term *anschaulich* is 'visualizable'. But, as in most languages, words that make reference to vision are not always intended literally. Seeing is widely used as a metaphor for understanding, especially for immediate understanding. Hence, *anschaulich* also means 'intelligible' or 'intuitive'.

Hilgevoord and Uffink (2011) note that the use of the term was in part a response to Schrodinger, whose claim was that his wave mechanics was more '*anschaulich*' (visualizable) than Heisenberg's matrix approach. Heisenberg's claim was that what is perceptible should be understood as what can be measured in experiments, rather than to our ability to mentally visualize a continuous causal path in the times in which measurements are not being conducted (which is closer to the sense in which Schrodinger used '*anschaulich*').

What Heisenberg referred to as the 'indeterminacy relation' soon came to be known as the 'uncertainty relation' or '(Heisenberg's) uncertainty principle'.

The simplest form of the principle can be stated as $\sigma_x \sigma_p \geq \hbar/2$ where $\sigma_x$ is the uncertainty in the position of a particle, $\sigma_p$ is the uncertainty in its momentum and $\hbar$ is the reduced Planck constant, $1.055 \times 10^{-34}$ J s. It places a lower bound on the precision to which the position and momentum of a particle can be simultaneously measured, and seems (based on all the empirical evidence available to date) to be a fundamental feature of the physical laws of the universe rather than an artefact of flaws in our own instruments and abilities to make measurements. Analysis can link a number of other complementary observables in similar ways.

Approaches to addressing the uncertainty principle experimentally in the more than 8 decades since it was described have been varied, but have typically involved the notion of preparing 'ensembles' of identical particles and conducting multiple measurements that are averaged to find the probability of particular states of position and momentum (or whichever other characteristics are being considered).

This approach to understanding the evolution of the variables through time and space is related to Born's statistical interpretation of the Schrodinger wave equation, which describes the quantum wavefunction. The wavefunction itself has to some extent been viewed as a mathematical artefact describing probabilities rather than something having an independent existence that can be accessed experimentally (Lundeen, et al., 2011).

**Two Recent Relevant Experiments**

Some recent work, published this year, has taken a different approach. It focuses on 'weak measurement' of one variable that do not significantly disturb the wavefunction followed by strong measurement of the other variable. It is still in a sense a statistical approach: the weak measurements do not give a precise value for the variable measured (this is the 'weakness' of the measurement) but a range of values, and these values are averaged over multiple trials to build up a picture of the wavefunction.

As Jeff Lundeen has noted in a comment on the online version of his paper in Nature (2011): "Weak measurement has suddenly become a hot topic this year." His team has published a paper in Nature, another team has published a paper in Science on which Sascha Kocsis is first author (2011) and a third team has published in Proceedings of the National Academy of Sciences (Goggin, et al., 2011)[1].

The Lundeen paper describes an experiment in which position was measured weakly followed by a strong measurement of momentum, while the Kocsis paper describes weak momentum measurements followed by strong position measurements, so together they offer a nice complementary (in a non-quantum sense) set of demonstrations of the possibilities.

This paper will discuss some of the history of the 'weak measurement' concept as well as the Lundeen et al. (2011) and Koscis et al. (2011) papers. It will consider the impact of weak measurement on physicists' understanding of the uncertainty principle, the meaning of the quantum wavefunction and the nature of measurement.

But first, a little more discussion of the uncertainty principle itself to prepare the ground.

---

[1] The Goggin et al. (2011) paper is beyond the scope of this particular discussion and will not be considered further, beyond noting it as evidence of growing interest in the field. The Leggett-Garg inequality in the title, however, is named in part for Leggett, who challenged weak measurement soon after it was first proposed (Leggett, 1989) and whose paper is briefly considered below.

**Busch, Heinonen and Lahti – positive statement of the uncertainty relations**

Paul Busch, Teiko Heinonen and Pekka Lahti wrote a relatively recent paper (2007) that explores some of the characteristics of the uncertainty principle. The authors note that it is usually considered as a *limitation* of what is possible, and is stated in
'negative' language such as 'it is impossible to...' They note that in fact it can have a very positive role as '...a condition ensuring that mutually exclusive experimental options can be reconciled if an appropriate trade-off is accepted' (p. 155).

Busch et al. (2007) focus on the specific case of the position-momentum variables as an illustration, and this is helpful in the present discussion because the Lundeen et al. (2011) and Koscis et al. (2011) papers also work with these observables. Particularly important is Busch et al.'s contention that '...it turns out that the extent of the disturbance of (say) momentum in a measurement of position can be controlled if a limitation in the accuracy of that measurement is accepted' (p. 155). This is the basis of the work on weak measurements. They continue:

> ...the uncertainty relation for measurement inaccuracies is not only a sufficient but also a necessary condition for the possibility of approximate joint measurements of position and momentum. (p. 156)

These authors survey attempts at experimental 'violations' of the uncertainty principle, and note that although claims have been made, particularly in relation to the experiment of Kim and Shih (1999), these have so far all been shown to actually be consistent with quantum mechanics and the uncertainty principle.

They clarify, however, that approximate (weak) measurements of the position (for example) can be conducted, followed sequentially by measurement of the perturbed momentum. The uncertainty principle will not be violated, but will instead provide a way of understanding and calculating the necessary 'trade-off' in the precision of the position measurement to yield a defined uncertainty in the momentum measurement (p. 173).

**A Brief History of Weak Measurement**

The concept of weak measurements is actually more than two decades old. Aharonov, Albert and Vaidman (AAV) (1988) considered the theoretical possibility of weak measurement experiments. Their approach is perhaps closer to what is described below as a 'quantum tomography' approach, however, in that it still relies on measurements over an ensemble of identically prepared particles, rather than the kind of averaging of measurements for individual particles, scanned across a range of values of the weakly-measured observable, used in the Lundeen et al. (2011) and Kocsis et al. (2011) experiments.

Leggett (1989), however, claimed that the concept of weak measurement outlined by Aharonov, Ablert and Vaidman (1988) was only of theoretical rather than practical interest based on considerations about the nature of measurement. His argument is a little beyond my pay grade at this point, however the subsequent work that has been done in the field suggests that his objections have either been shown not to be substantive or that the issues he raises about the nature of measurement are considered not to be fatal for the approach. Peres (1989) also challenged some features of the AAV proposal.

Aharonov and Vaidman (1989) responded, and defended their concept against Leggett's (1989) and Peres' (1989) critiques by clarifying some points about their model in relation to the issues raised. The issues seem to be in some ways more philosophical than procedural, however, in terms of what constitutes 'measurement'.

Ritchie, Story and Hulet (1991), in fact, experimentally demonstrated weak measurement 20 years ago, only 2 years after the Aharonov, Albert and Vaidman (1988) paper appeared. While the AAV theoretical treatment used spin-½ particles to illustrate the concept, Ritchie, Story and Hulet used photons (from a laser) which are spin-1. Their work uses polarisation for the measurement, a small value of the polarisation angle α and a number of other features used in the current Lundeen et al. (2011) and Kocsis et al. (2011) experiments.

More recently, Ahshab and Nori (2009) have shown how the results achieved with weak measurements can be achieved in an apparatus not incorporating weak measurements and also considered what they call "the correct physical description, according to quantum mechanics, of what is being measured in a weak-value-type experiment" (p. 1). They claim that it is not appropriate to consider the weak measurement of one observable followed by the strong measurement (through post-selection) of a complementary observable as two separate measurement events, but rather consider that quantum-mechanically they ought to be considered as a single combined-measurement situation.

**Lundeen et al. – weak measurement of photon position**

Jeff Lundeen and team at the Canadian Institute for National Measurement Standards commented:

> The wavefunction is the complex distribution used to completely describe a quantum system, and is central to quantum theory. But despite its fundamental role, it is typically introduced as an abstract element of the theory with no explicit definition. Rather, physicists come to a working understanding of the wavefunction through its use to calculate measurement outcome probabilities by way of the Born rule. At present, the wavefunction is determined through tomographic methods, which estimate the wavefunction most consistent with a diverse collection of measurements. The indirectness of these methods compounds the problem of defining the wavefunction. Here we show that the wavefunction can be measured directly by the sequential measurement of two complementary variables of the system. The crux of our method is that the first measurement is performed in a gentle way through weak measurement, so as not to invalidate the second. The result is that the real and imaginary components of the wavefunction appear directly on our measurement apparatus. (2011, p. 188)

Weak measurement, as Busch et al. (2007) noted, trades precision in the measurement of one observable for reduced perturbation in the complementary variable measured at a later time. The weak measurement yields a range of possible values for the first variable rather than a specific value.

Lundeen et al. (2011) note that the common approach to measuring the wavefunction can be described as 'quantum tomography': combining large numbers of distinct measurements of ensembles of identically-prepared systems and estimating the wavefunction most compatible with these measurements. They state that their own method allows more direct measurement of the wavefunction: indeed, that the imaginary and real components of the wavefunction 'appear directly on our measurement apparatus' (p. 188).

The question of 'measurement' is a vexed one in quantum mechanics – what constitutes a 'measurement'? Plenty has been written on the topic (Busch et al. (1996) provide a good overview) and in general it reduces to the notion that a 'measurement' moves some 'pointer'. A weak measurement occurs when the strength of the coupling between the physical system being measured and the pointer is reduced. The weaker coupling reduces the induced perturbation in the quantity being measured. It also reduces the precision of the measurement, but averaging many measurements can compensate for this reduced precision. As the strength of the coupling approaches zero, the perturbation also approaches zero... but the precision of the measurement decreases.

The weak measurement does not significantly affect the result gained in a **subsequent** measurement of a complementary (i.e. uncertainty principle limited) observable. In the Lundeen et al. (2011) experiment, this subsequent measurement was of momentum and was used to 'post-select' photons with particular characteristics – in this instance, zero transverse momentum.

Photons (either single photons emitted by 'spontaneous parametric down-conversion' or an attenuated beam from a laser) were introduced to the apparatus using a single mode optical fibre. They were polarised using a microwire polariser then collimated with a lens. The resulting beam was 'masked' through a rectangular aperture. A sliver of a half-wave ($\lambda/2$) plate was used to weakly measure the position of the photons 45 mm after the lens. A Fourier transform lens was then used in the beam, and photons that had a value of p=0 for the transverse momentum 'post-selected' by passing the beam through a 15 μm slit on the axis. The beam was then collimated with yet another lens and passed through either a half-wave ($\lambda/2$) or a quarter-wave ($\lambda/4$) plate and then through a polarising beam-splitter. The difference between the readings at the two detectors represents the difference between the real (λ/2) and imaginary ((λ/4) parts of the wavefunction.

The keys to this experiment are (1) the use of post-selection of photons with zero transverse momentum, so that the momentum for the wavefunction is fixed and known and (2) the use of the polarisation of the photons as the 'pointer' for the measurement. The linear transverse polarisation is proportional to the real part of the wavefunction and the circular polarisation to the imaginary part. This second facet requires more explanation.

For an arbitrary variable A which is complementary with another arbitrary variable C, the average of the weak measurement will have the value <Ψ|A|Ψ> which will be represented by a movement of the pointer proportional to this value. For the subset of the ensemble where C yields the result c (e.g. in this experiment, where p yields the value 0), an expression for the „weak expectation value" in the limit of zero coupling strength can be given by:

$$\langle A \rangle_w = \frac{\langle c|A|\Psi \rangle}{\langle c|\Psi \rangle}$$

Lundeen et al. (2011) note that:

> Unlike the standard expectation value, <A>, the weak value <A>$_w$ can be a complex number. This seemingly strange result can be shown to have a simple physical manifestation: the pointer's position is shifted by Re<A>$_w$ and receives a momentum kick of Im<A>$_w$. The complex nature of the weak value suggests it could be used to indicate both the real and the imaginary parts of the wavefunction (p. 188).

This is what they have in fact demonstrated. For a photon, the weak measurement of the position, $\pi_x$, is followed by a strong measurement of the momentum, yielding the result P = p. The weak expectation value for the position is given by:

$$\langle \pi_x \rangle_w = \frac{\langle p|x \rangle \langle x|\Psi \rangle}{\langle p|\Psi \rangle} = \frac{e^{ipx/\hbar} \Psi(x)}{\Phi(p)}$$

Given that the experiment post-selects for photons with p=0, this simplifies to:

$$\langle \pi_x \rangle_w = k \Psi(x)$$

where:

$$k = \frac{1}{\Phi(0)}$$

The k term is a constant that can be factored out by normalising the wavefunction.

The average value of the weak measurement, $\pi_x$, is proportional to the wavefunction at x, and scanning through x allows the complete wavefunction to be described. "At each x, the observed position and momentum shifts of the measurement pointer are proportional to Re$<\Psi>_w$ and Im$<\Psi>_w$ respectively" (Lundeen et al., 2011, p. 188).

The weak measurement of position is accomplished by changing the polarisation of the photons through an angle α. If α= 90º this would constitute a strong measurement, since it would be possible to perfectly discriminate between orthogonal polarisations. Reducing the angle α reduces the strength (and therefore the precision) of the measurement, and control of the angle allows it to be reduced to an arbitrarily weak measurement. This is exactly the 'trade-off' – informed by, and indeed made possible to calculate based on, the uncertainty principle – described by Busch et al. (2007).

If the initial polarisation is treated as a spin-½ spin-down vector, the weak value for the position is given by the expression:

$$<\pi_x>_w = \frac{1}{\sin\alpha}\left(<s|\sigma_x|s> - i<s|\sigma_y|s>\right)$$

where $\sigma_x$ is the Pauli x matrix and $\sigma_y$ is the Pauli y matrix and "s is the final polarisation state of the pointer" (Lundeen, 2011, p. 189). The expectation values for $\sigma_x$ is found by sending the photon through a half-wave plate and for $\sigma_y$ using a quarter-wave plate and then using a polarising beam splitter to measure the difference between the two polarisations.

The team set α=20º and scanned through the values of x in 1 mm increments. They also tested the approach using a number of different test wavelengths.

Lundeen et al. (2011) conclude:

> This is the general theoretical result of this Letter. It shows that in any physical system one can directly measure the quantum state of that system by scanning a weak measurement through a basis and appropriately post-selecting in the complementary basis. (p. 190)

This issue of 'direct' measurement of the wavefunction is discussed further below.

**Kocsis et al. – weak measurement of photon momentum**

The Kocsis et al. (2011) team is very international, consisting of scientists based in Canada, Australia (including Griffith University), France and the USA.

This team also works in the domain of position and momentum, but is concerned with the old problem of which path a photon takes through a double-slit interferometer. They note that Feynman (1989) considered that the double-slit experiment contains "the heart of quantum mechanics". Quantum theory and the uncertainty principle have traditionally been held to prohibit concurrent knowledge of the path of a photon (i.e. which slit it passed through) and interference fringes.

The Kocsis et al. experiment is constituted by the following steps:

- Single photons from a quantum dot are split using a 50:50 beam splitter.

- Quarter-wave plates and half-wave plates allow the number of photons passing through each 'slit' to be varied.

- They are then passed through a polariser in order to prepare them in a diagonal polarisation characterised as $|D\rangle = 1/\sqrt{2}(|H\rangle+|V\rangle)$ where $|H\rangle$ is identified with the x axis and $|V\rangle$ with the y axis.

- The photons are then reflected in parallel, in such a way that a double slit is simulated.

- The photons undergo a weak measurement of their momentum using a 0.7 mm thick piece of calcite with its optical axis at $42^o$ in the x-y plane. This rotates the polarisation state. A quarter-wave plate and displacer is used to extract the weak measurement of k.

- Lenses allow different imaging planes to be measured.

- The final position of the photons is measured (strongly) using a charge-coupled device (CCD).

As noted earlier, the Kocsis et al. (2011) experiment offers a nice counterpoint to the Lundeen et al. (2011) work because it measures momentum weakly followed by a strong measurement of position, while the Lundeen team's approach is the opposite.

The polarisation degree of freedom of the photons was used as a pointer to weakly measure the momentum of the photons. The combination of the weak and strong measurements does not allow the measurement of the trajectory of an individual photon through the double-slit apparatus: as Kocsis et al. (2011) note, such a concept is not even well defined for quantum particles. It does, however, allow the calculation of average trajectories. Here is how the authors summarise the contribution made by their work:

> Our experimentally observed trajectories provide an intuitive picture of the way in which a single particle interferes with itself. It is of course impossible to rigorously discuss the trajectory of an individual particle, but in a well-defined operational sense we gain information about the average momentum of the particle at each position... (p. 1173)

**Discussion**

It is important to note that neither of these papers makes the claim that the uncertainty relations have been violated, or even really circumvented. The use of weak measurements of one observable combined with strong measurement of a complementary observable exploits the 'trade-off' described by Busch et al. (2007), and occurs within the limitations imposed by the uncertainty relations.

What, then, is the scientific significance of this 'hot field'? Lundeen et al. (2011) make a large claim for the significance of their work:

> In our direct measurement method, the wavefunction manifests itself as shifts of the pointer of the measurement apparatus. In this sense, the method provides a simple and unambiguous operational definition of the quantum state: it is the average result of a weak measurement of a variable followed by a strong measurement of the complementary variable. (p. 190)

This is a large claim in the sense that it is an attempt to supplant the time-honoured understanding of the wavefunction in terms of Born's statistical interpretation – as a mathematical system for calculating the probabilities of measuring particular defined eigenfunctions of a system. Lundeen et al. (2011) are claiming that the wavefunction itself has some form of measurable, physical reality, and that both its real and imaginary elements can be measured.

The measurement is still not 'direct', in that it involves averaging of a large number of measurements of the weakly measured observable along with post-selection of the strongly measured observable. Essentially, the weak measurement does not collapse the wavefunction to a single eigenstate, but only to a superposition of a number of possible eigenstates. While strong measurements must be real, weak measurements can be complex.

While the reported empirical results are compelling, I suspect that much of the controversy (to the extent that any remains) is in matters of definition rather than more substantive issues.

Work reconciling the experimental results reported by Lundeen et al. (2011) and Kocsis et al. (2011) (and the results from Goggins et al. (2011), not discussed in this paper) with Ahshab and Nori's (2009) suggestion that the measurements should be considered separately rather than together would also be potentially very interesting.

**Conclusion**

While it has its origins over 20 years ago, the publication of three papers on weak measurements in 2011 suggests that there is interesting work being done and remaining to be done. A slightly different approach to understanding the physical meaning of the wavefunction is one very interesting consequence of this work. There remain controversies about the meaning of 'measurement' and what the measurements gained in weak measurement experiments really 'mean' from a quantum mechanical perspective, meaning there is interesting and important theoretical as well as experimental work remaining to be done in this field.

# References

Aharonov, Y., Albert, D.Z. & Vaidman, L. (1988). How the result of the measurement of a component of the spin of a spin-½ particle can turn out to be 100. *Phys Rev Lett, 60*, 1351- 1354. DOI: 10.1103/PhysRevLett.60.1351

Aharonov, Y. & Vaidman, L. (1989). Reply to Leggett and Peres. *Phys Rev Lett, 62*, 2327. DOI: 10.1103/PhysRevLett.62.2327

Ahshab, S. & Nori, F. (2009). How the measurement of a component of the spin of a spin-½ particle can turn out to be 100 without using weak measurements. Measurement. arXiv: 0907.4823

Busch, P., Heinonen, T. & Lahti, P. (2007). Heisenberg's uncertainty principle. *Physics Reports, 452*(6): 155-176. DOI:10.1016/j.physrep.2007.05.006.

Busch, P., Lahti, P.J. & Mittelstaedt, P. (1996). *The quantum theory of measurement.* Dordrecht: Springer.

Feynman, R. (1989). *The Feynman Lectures on Physics*. Boston: Addison-Wesley.

Gao, S. (2011). Meaning of the wave function. *International Journal of Quantum Chemistry*. 10.1002/qua.22972. Forthcoming paper viewed in online preview. Accessed 18/9/11. http://onlinelibrary.wiley.com.ezproxy.library.uq.edu.au/documentcitationdownload?publicationDoi=10.1002/%28ISSN%291097-461X&doi=10.1002/qua.22972&type=journal

Goggin, M.E., Almeida, M.P., Barbieri, M., Lanyon, B.P., O'Brien, J.L., White, A.G. & Pryde, G.J. (2011). Violation of the Leggett–Garg inequality with weak measurements of photons. *PNAS January 25, 2011 vol. 108* no. 4 1256-1261. DOI: 10.1073/pnas.1005774108.

Heisenberg, W. (1927) „Ueber den anschaulichen Inhalt der quantentheoretischen Kinematik and Mechanik" *Zeitschrift für Physik* **43** 172-198. English translation in Wheeler, J.A. and Zurek, W.H. (eds) (1983) *Quantum Theory and Measurement* (Princeton NJ: Princeton University Press), pp. 62-84.

Hilgevoord, Jan and Uffink, Jos, (2011). The Uncertainty Principle, *The Stanford Encyclopedia of Philosophy* (Spring 2011 Edition), Edward N. Zalta (ed.). Accessed 18/9/11. http://plato.stanford.edu/archives/spr2011/entries/qt-uncertainty/

Kim, Y.-H. & Shih, Y.H. (1999) Experimental realization of Popper's experiment: violation of the uncertainty principle? *Found. Phys. 29:* 1849–1861.

Kocsis, S., Braverman, B., Ravets, S., Stevens, M.J., Mirin, R.P., Shalm, L.K. & Steinberg, A.M. (2011). Observing the Average Trajectories of Single Photons in a Two-Slit Interferometer. *Science, 332*(6034): 1170-1173. DOI: 10.1126/science.1202218.

Leggett, A.J. (1989). Comment on „‟„How the result of a measurement of a component of the spin of a spin-1/2 particle can turn out to be 100‟‟‟, *Phys. Rev. Lett. 62*, 2325. DOI: 10.1103/PhysRevLett.62.2325

Lundeen, J.S., Sutherland, B., Patel, A., Stewart, C. & Bamber, C. (2011). Direct measurement of the quantum wavefunction. *Nature, 474*(7350), 188-191. DOI:10.1038/nature10120.

Peres, A. (1989). Quantum measurements with post-selection. *Phys Rev Lett, 62*, 2326. DOI: 10.1103/PhysRevLett.62.2326.

Ritchie, N.W.M., Story, J.G. & Hulet, R.G. (1991). Realization of a measurement of a „weak value‟. *Phys Rev Lett, 66*, 1107-1110. DOI: 10.1103/PhysRevLett.66.1107

Wheeler, J.A. and Zurek, W.H. (eds) (1983) *Quantum Theory and Measurement* (Princeton NJ: Princeton University Press), pp. 62-84.